\documentclass[aps,prb,twocolumn,showpacs,superscriptaddress]{revtex4}
\usepackage{graphicx}
\usepackage{amsfonts,amssymb,amsmath}
\usepackage{bm}
\begin{document}

\title{A significant influence of the substrate on the magnetic 
anisotropy of monatomic nanowires}

\date{\today}
 
\author{Matej Komelj}
\email{matej.komelj@ijs.si}
\affiliation{Jo\v zef Stefan Institute, Jamova 39, SI-1000 Ljubljana,
  Slovenia}
\author{Daniel Steiauf}
\affiliation{Max-Planck-Institut f\" ur Metallforschung, Heisenbergstra{\ss}e 3,
D-70569 Stuttgart, Germany}
\author{Manfred F\" ahnle}
\affiliation{Max-Planck-Institut f\" ur Metallforschung, Heisenbergstra{\ss}e 3,
D-70569 Stuttgart, Germany}
\begin{abstract}
The magnetic anisotropy energy of Fe and Co 
monatomic nanowires both free-standing and at the step edge of a Pt surface is investigated within the framework of the 
density-functional theory and local-spin-density (LSDA) approximation. 
Various types of the analysis of the calculated data reveal that the 
spin-orbit interaction of the Pt atoms and the hybridization 
between the electronic states
have an important impact on the direction of the easy axis and on the magnitude
of the magnetic anisotropy, both by a direct contribution localized at the 
Pt atoms and by an indirect contribution due to the modification of the 
Co-localized part via hybridization effects.
\end{abstract}
\pacs{75.75.+a, 75.30.Gw, 71.15.Mb}
\maketitle
\section{Introduction}
The experimentally observed existence of ferromagnetism in linear chains
of Co atoms, grown at a step edge of a Pt(997) surface 
terrace\cite{Gambardella:2002}, motivated several 
theoretical investigations which were aimed to interpret the measured large
orbital contribution to the magnetic moments\cite{Komelj:2002,Ederer:2003},
and the magnetic anisotropy\cite{Hong:2003,Lazarovits:2003,Ujfalussy:2004,
Shick:2004,Shick:2005,Maca:2006}. The latter is distinguished by the
experimentally found easy axis along the direction perpendicular to the Co
wire, shifted by $43^\circ$ from the surface normal towards the Pt step edge, 
and by the 
magnetic-anisotropy energy (MAE) of the order of $2\>{\rm meV/Co\>atom}$. 
This value is a factor of about 50 larger than the one of hcp Co (which 
is already large for a transitional metal).
Various theoretical attempts to reproduce the experimental findings yielded
quantitatively different results. Hong and Wu\cite{Hong:2003}  performed the 
full-potential linearized-augmented-plane-waves  (FLAPW) calculation for
infinite Co wires on the flat Pt(001) and Cu(001) substrates. They found
that the MAE magnitude was enhanced when they replaced the Cu substrate by 
platinum due to the strong spin-orbit coupling (SOC) of Pt atoms. 
They obtained the easy axis along the wire in the case of the Cu(001) substrate,
and perpendicular to the wire  in the plane of the Pt(001) substrate. 
They also noticed a certain sensitivity of the results on the size of the 
supercell.  The results of the Korringa-Kohn-Rostoker (KKR) 
calculations\cite{Lazarovits:2003,Ujfalussy:2004} on finite (up to 10 
atoms) Co wires along a step edge of a Pt(111) surface were in remarkable agreement with the experiment.
However, to some 
extent, this agreement might be coincidental since the atomic-sphere 
approximation (ASA) was applied for the potential. Again, the authors ascribed
the peculiar magnetic anisotropy to the influence of the Pt substrate. 
The same conclusion was drawn also on the basis of the FLAPW calculations 
for the infinite Co\cite{Shick:2004,Shick:2005,Maca:2006} and 
Fe\cite{Shick:2005} wires at the Pt(111) step edge by analyzing the data
in the frame of a symmetry-based ``atomic-pair'' model. Although shape
approximations were not applied for the potential, the experimental easy axis
was not reproduced as good as in the case of the KKR ASA 
calculation\cite{Lazarovits:2003,Ujfalussy:2004}, even if the structural
relaxations were taken into account\cite{Maca:2006}. 
Finally, an FLAPW calculation\cite{Bihlmayer:2005} for a considerably larger
supercell as used in Ref. \onlinecite{Maca:2006}, including structural 
relaxations,
found an easy axis which is in a plane parallel to the substrate surface and
almost perpendicular to the wire, in contrast to the result of Ref. 
\onlinecite{Maca:2006} and to the experimental result. All this indicates that
the orientation of the calculated easy axis depends extremely sensitively 
on the calculational details. We therefore do not take the enormous effort
to converge the results with respect to all parameters of the calculation
(e.g., geometry and size of the supercell), but we concentrate on the physical
interpretation of the results concerning the influence of the Pt substrate by
means of two different types of the analysis. 
\section{Calculational method}
We performed the calculations with the FLAPW\cite{Wimmer:1981} 
Wien97\cite{Blaha:1990} code by using the local-spin-density approximation
(LSDA)\cite{Perdew:1992-1} for the exchange-correlation potential.  
The SOC contribution was added to the Hamiltonian in terms of the 
second-variational method\cite{Singh:1994,Novak?}. In the implementation of
Ref. \onlinecite{Blaha:1990} 
SOC is taken into account just within muffin-tin spheres  but not in the 
interstitial region.
The energy difference
in the total energy due to different magnetization directions was determined
according to the magnetic force theorem\cite{Mackintosh:1980,Weinert:1985}.
In this approximation, first a self-consistent effective potential
$V_{\rm eff}^{(0)}$ is determined for an initial magnetic configuration (0) 
of the system. In a subsequent step the eigenvalues are determined for 
respectively two different orientations (1) and (2) of the magnetization by 
respectively one single diagonalization of the Hamiltonian for the same
fixed potential $V_{\rm eff}^{(0)}$ but with the SOC term corresponding to 
the two orientations of the magnetization. The magnetic anisotropy energy
$e_{\rm MCA}$ which is defined as the difference in the total energy for the 
two magnetization directions is then approximated by the difference in the 
respective sums of eigenvalues:
\begin{equation}
e_{\rm MCA}^{\rm ft}=
\int_{-\infty}^{E_{\rm F}^{(2)}}\epsilon{\cal N}^{(2)}
\left(\epsilon\right)d\epsilon-
\int_{-\infty}^{E_{\rm F}^{(1)}}\epsilon{\cal N}^{(1)}
\left(\epsilon\right)d\epsilon,
\end{equation}
In eq. (1) the functions ${\cal N}^{(1)}\left(\epsilon\right)$ and 
${\cal N}^{(2)}\left(\epsilon\right)$  are the electronic density of states
for the two orientations, and $E_{\rm F}^{(1)}$ and $E_{\rm F}^{(2)}$ are
the respective Fermi levels, which are calculated from the number $Z$ of the valence electrons according to:
\begin{equation}
Z=\int_{-\infty}^{E_{\rm F}^{(i)}}{\cal N}^{(i)}\left(\epsilon\right)d\epsilon.
\end{equation}
The convergence of the 
magnetic anisotropy energy with respect to the number of $k$ points used
for the sampling of the Brillouin zone is notoriously bad. However, it turns
out that less $k$ points are needed for the calculation of $V_{\rm eff}^{(0)}$.
Hence, the advantage of the magnetic force theorem is that the calculation 
of $e_{\rm MCA}^{\rm ft}$ from (1) requires considerably less computational 
power than the determination of the respective difference in the total energy, 
obtained from two self-consistent calculations. 
Because the FLAPW method is 
very time consuming, the application of the force theorem is desirable.
An open question thereby is how to select the initial configuration 
(0) for the determination of $V_{\rm eff}^{(0)}$ in order to minimize  
the difference between $e_{\rm MCA}$ and $e_{\rm MCA}^{\rm ft}$. To figure this
out we have calculated by means of the linear-muffin-tin-orbital (LMTO)
method in atomic-sphere approximation \cite{Andersen:1975} both
$e_{\rm MCA}$ (which is possible because LMTO-ASA is much faster, albeit
less accurate, than FLAPW) and $e_{\rm MCA}^{\rm ft}$ for various
configurations (0). It turns out that it is absolutely indispensable to 
include SOC for the initial configuration. The initial orientation of the 
magnetization has only a subtle effect on $e_{\rm MCA}^{\rm ft}$ since for
each orientation the agreement between $e_{\rm MCA}$ and $e_{\rm MCA}^{\rm ft}$
is nearly perfect as presented in Fig. 1.  For this test, the infinite Co 
\begin{figure}
\includegraphics[scale=0.3]{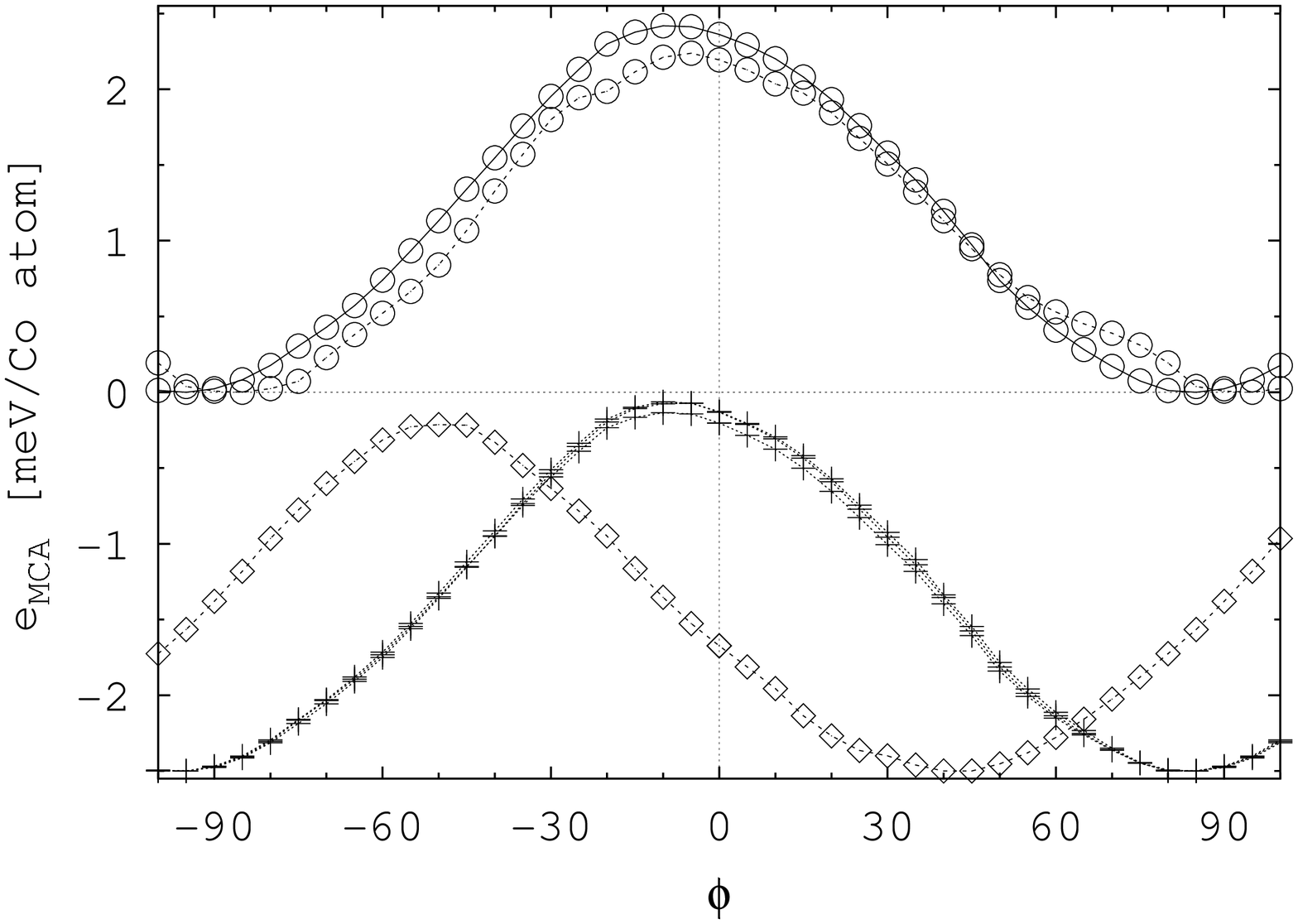}
\caption{Comparison between total energy differences $e_\mathrm{MCA}$ and force theorem results $e^\mathrm{ft}_\mathrm{MCA}$ from LMTO-ASA calculations. The total energy (symbol $\small\bigcirc$) was obtained with $20\times 4\times 1$ (dashed  line) and $32\times 6\times 1$ (solid line) $k$ points for the sampling of the Brillouin zone. All the force-theorem calculations were done with $32\times 6\times 1$ $k$ points, starting from a potential $V_\mathrm{eff}^{(0)}$ created with $20\times 4\times 1$ $k$ points. The $\Diamond$ starting potential was without SOC, while the $+$ starting potentials included SOC with the magnetization in different directions ($\phi=0^\circ, \pm 90^\circ, 180^\circ$ with $\theta=0^\circ$; and $\theta=90^\circ$ with $\phi=0^\circ$).
}
\end{figure}
nanowire at the step edge of the Pt(111) surface was modelled 
by a supercell with the geometry from Ref. \onlinecite{Shick:2004}, which 
contains 16 atoms 
(15 Pt + one Co atom) and 8 empty atomic spheres that are required due to the 
ASA approximation. 
\par
The rest of the calculations were performed for a monatomic wire at the step edge of a Pt surface described by a supercell of 13 atoms (12 Pt + one Co or Fe atom) by adopting the experimental lattice
parameters of fcc Pt (for details, see Ref. 
\onlinecite{Komelj:2002}). In addition, freestanding wires were considered by removing
all Pt atoms from the supercell.  
The orientation of the coordinate system was adopted from 
Refs.\onlinecite{Shick:2004,Shick:2005,Maca:2006} (see the inset in Fig. 2).
\begin{figure}
\includegraphics[scale=0.4,angle=0]{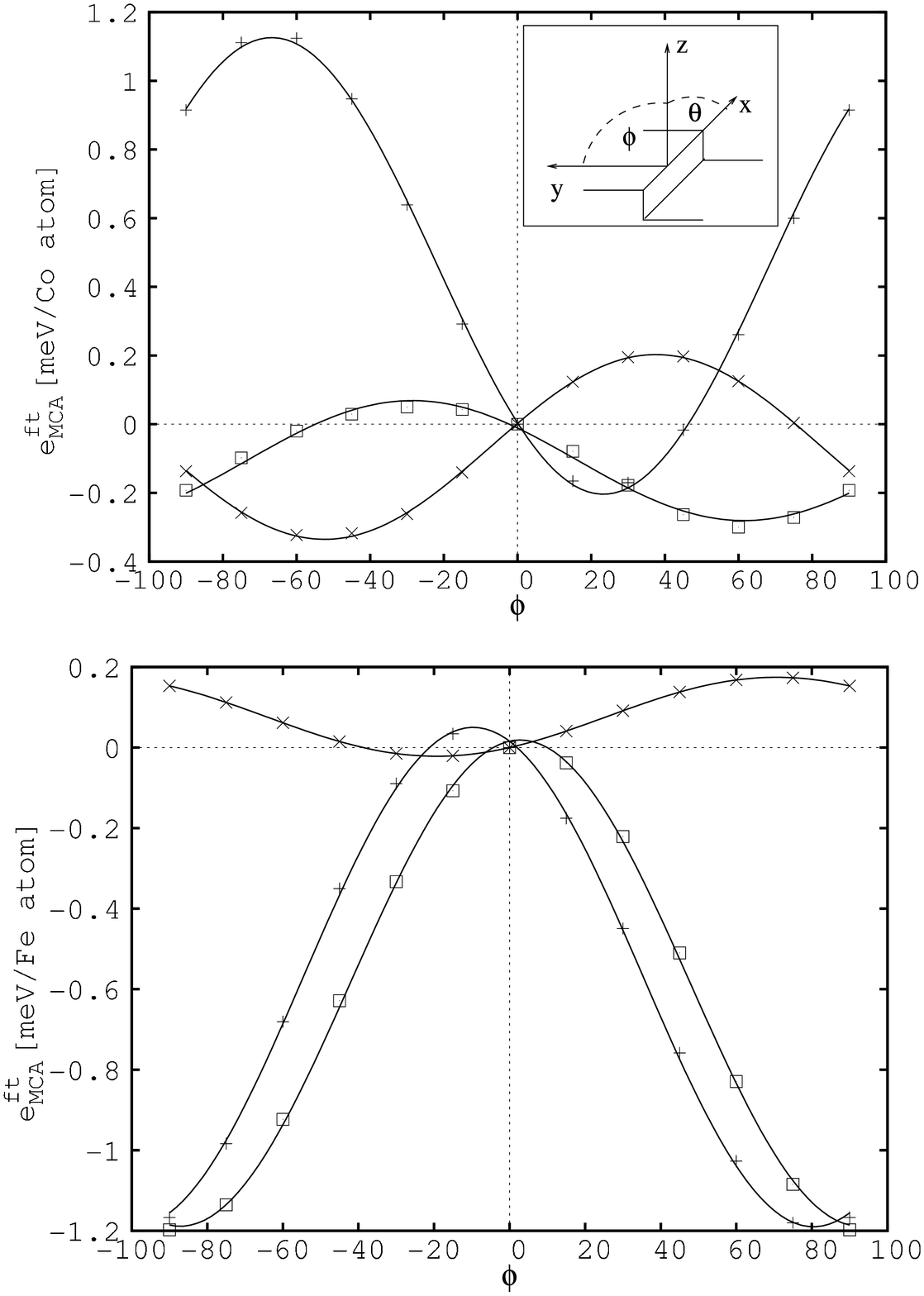}
\caption{The calculated energy difference
$e^{\rm ft}_{\rm MCA}=e(\theta=0^\circ,\phi)-e(\theta=0^\circ,\phi=0^\circ)$
for the Co (top) and the Fe (bottom) Pt-supported wire with SOC at all
sites $+$ or just at the Co(Fe)  $\times$ or Pt $\square$ atoms.}
\end{figure}
The cut-off parameter for the plane-wave 
expansion was set to $7.3\>{\rm Ry}$.
The Brillouin-zone (BZ) integration was 
carried out by both the modified tetrahedron\cite{Blochl:1994} and
the Gaussian smearing\cite{Fu:1983} methods 
with 45 (40 in the case of the freestanding wires) k-points in 
the full BZ for the self-consistent part and 192 (184) for the force-theorem
part of the calculation as determined on the basis of convergence tests.
\par
We calculated $e_{\rm MCA}^{\rm ft}$ for the magnetization 
confined
either to the $yz$ (the angle $\phi$, Fig. 2) or the $xz$ plane 
(the angle $\theta$, Fig. 3) 
\begin{figure}
\includegraphics[scale=0.4,angle=0]{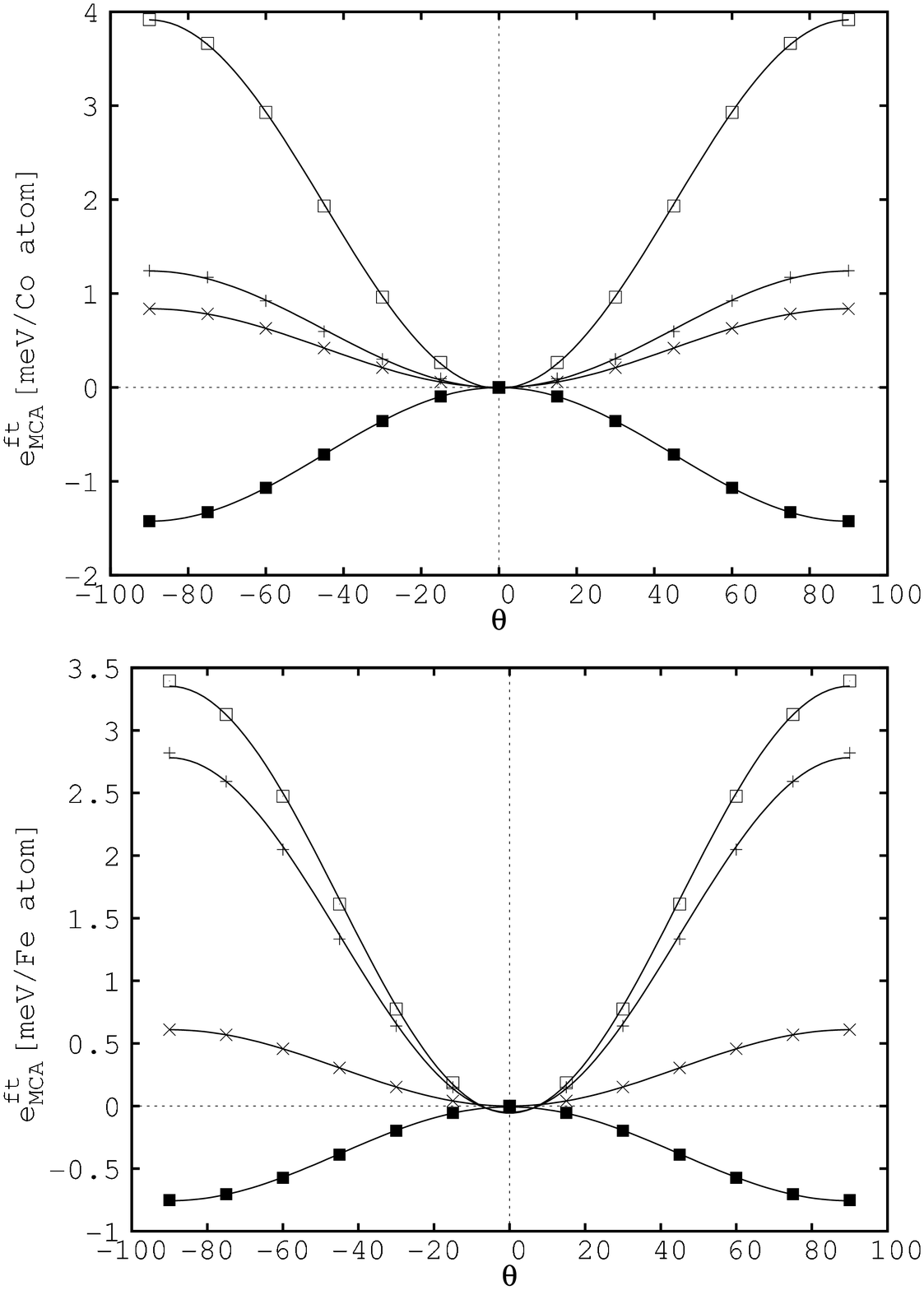}
\caption{The calculated energy difference
$e^{\rm ft}_{\rm MCA}=e(\theta,\phi=0^\circ)-e(\theta=0^\circ,\phi=0^\circ)$
for the Co (top) and the Fe (bottom) Pt-supported wire with SOC at all
sites $+$ or just at the Co(Fe)  $\times$ or Pt $\square$ atoms. The symbols
$\blacksquare$ present the respective energy difference for the freestanding
wires.}
\end{figure}
with respect to the energy for the case of the magnetization along
the $z$ axis. The calculated data were fitted by the functions $a-b\cos^2\left(
\phi-c\right)$ and $a-b\cos^2\left(\theta\right)$ for the respective 
planes. 
Since there is a quantitative deviation from the results obtained by 
Shick {\it et al.}\cite{Shick:2004,Shick:2005} for a slightly different 
supercell, we performed a test calculation
with their type of the supercell (16 atoms but considerably less vacuum
at the  top of the (111) surface), and we could satisfactorily reproduce the 
published 
results from Refs.\ \onlinecite{Shick:2004,Shick:2005}. This demonstrates that 
different computer codes give the same results, and it confirms the findings 
from
previous investigations that the magnetic anisotropy depends extremely 
sensitively on the details of the structural model. 
\par
To explore the influence of the Pt substrate we performed two types of the
analysis:
\par\noindent
(a) The densities of states ${\cal N}^{(1,2)}\left(\epsilon\right)$ in eq. (1)
can be subdivided into the contributions 
${\cal N}^{(1,2)}_j\left(\epsilon\right)$  which correspond to the respective
projections onto the Co (Fe) atoms ($j=1$) and Pt atoms ($j=2$) muffin-tin
spheres. This subdivision yields $e_{\rm MCA}^{\rm ft,Co}$ 
($e_{\rm MCA}^{\rm ft,Fe}$) and $e_{\rm MCA}^{\rm ft,Pt}$. Because
the total $e_{\rm MCA}^{\rm ft}$ contains also the contribution from the
interstitial region between the muffin-tin spheres, we cannot expect that
the sum $e_{\rm MCA}^{\rm ft,Co}+e_{\rm MCA}^{\rm ft,Pt}$ is equal to
$e_{\rm MCA}^{\rm ft}$. In the following we call $e_{\rm MCA}^{\rm ft,Pt}$ the
``direct contribution'' of the Pt atoms to the magnetic anisotropy energy,
because it is directly assigned to the Pt sites. Within the framework
of the rigid band model we can determine these quantities as a function of the 
band filling $Z+\Delta Z$, where $Z$ denotes the band filling for the real system.
The change in the band filling leads to
the shifts in the Fermi levels appearing in eq. (1) as:
$Z+\Delta Z=\int_{-\infty}^{E_{\rm F}^{(1,2)}+\Delta E_{\rm F}^{(2,1)}}
{\cal N}^{(1,2)}\left(\epsilon\right)d\epsilon$. The results of this
analysis are presented in Figs.\ 4,5.\par\noindent
\begin{figure}
\includegraphics[scale=0.4,angle=0]{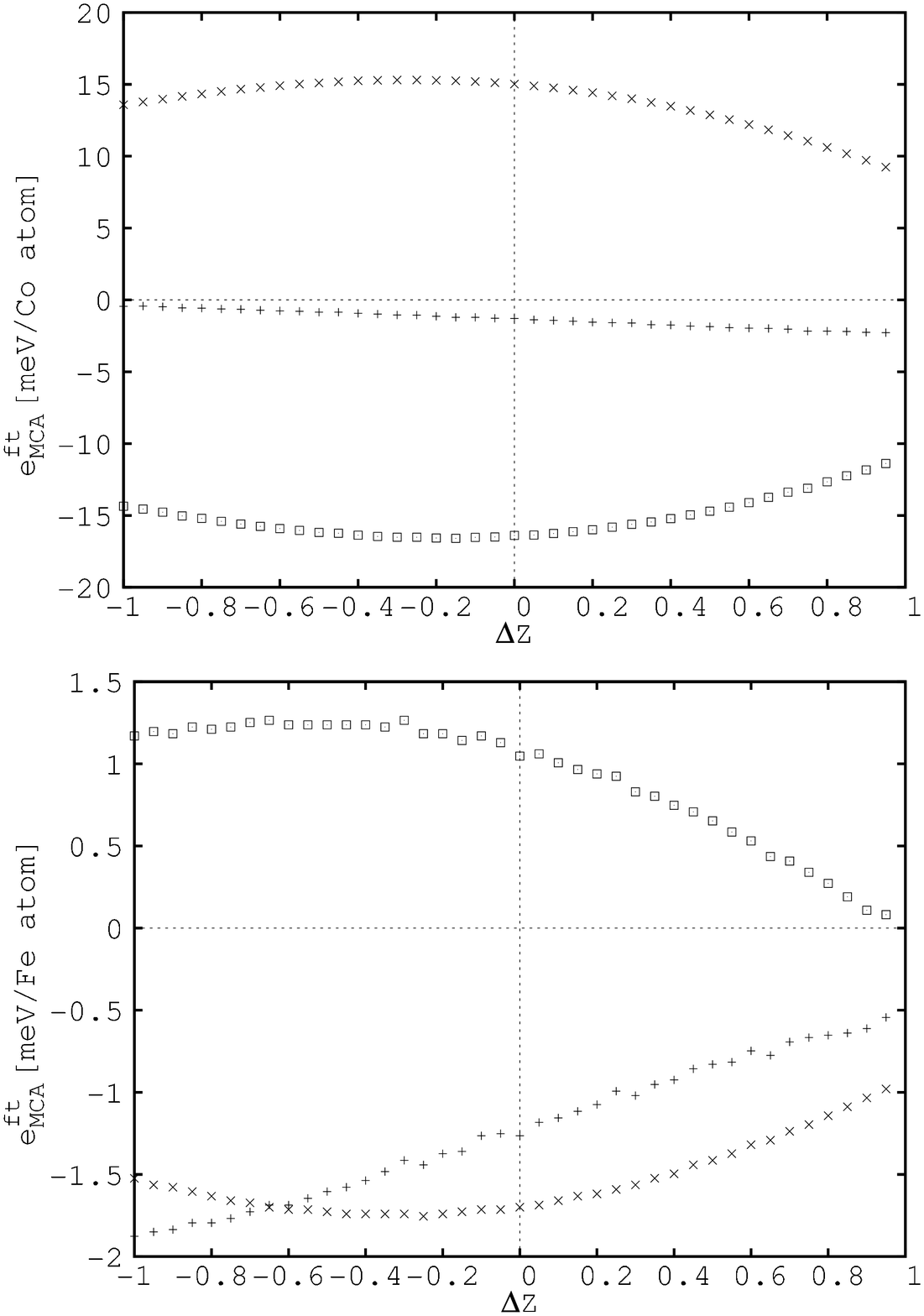}
\caption{The calculated magnetic anisotropy energy
$e_{\rm MAE}^{\rm ft}=e(\theta=0^\circ,\phi=22.85^\circ)-
e(\theta=0^\circ,\phi=-67.15^\circ)$
for the Co wire (top) and
$e_{\rm MAE}=e(\theta=0^\circ,\phi=80.40^\circ)-
e(\theta=0^\circ,\phi=-9.60^\circ)$
for the Fe wire (bottom); total contribution ($+$), Co (Fe)
contribution ($\times$), Pt contribution ($\square$).}
\end{figure}
\begin{figure}
\includegraphics[scale=0.4,angle=0]{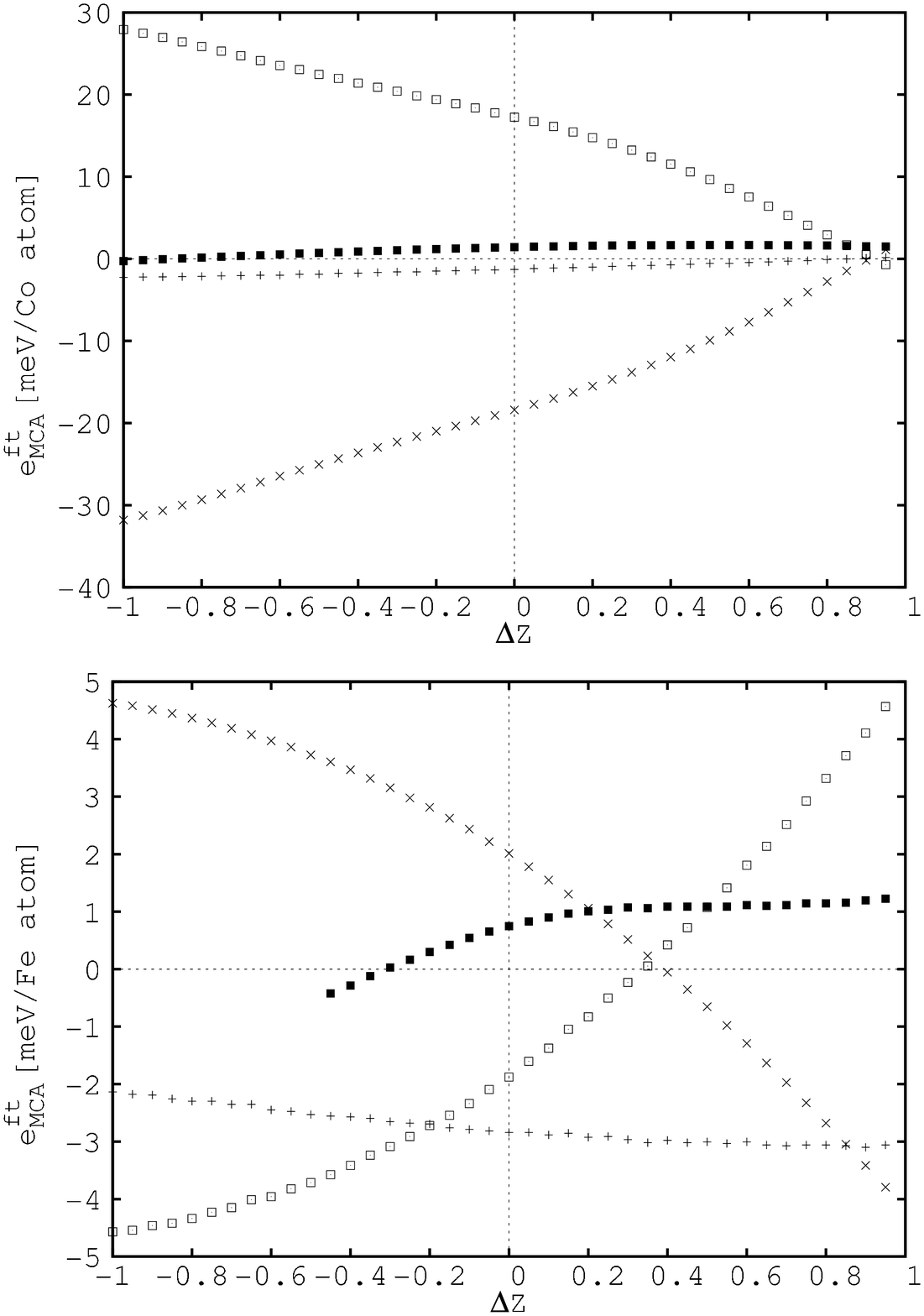}
\caption{The calculated magnetic anisotropy energy
$e_{\rm MAE}=e(\theta=0^\circ,\phi=0^\circ)-
e(\theta=90^\circ,\phi=0^\circ)$ for the Co (top) and the Fe (bottom) wire;
total contribution ($+$), Co (Fe)
contribution ($\times$), Pt contribution $\square$, free-standing
wire ($\blacksquare$).}
\end{figure}
(b) In addition to the calculation with SOC for all atoms, we performed 
calculations where we switched off the SOC term for the Co(Fe) atoms or 
for the Pt atoms, respectively.  Thereby we changed the hybridization 
between the electronic states at various sites and hence the eigenvalues 
for the two orientations. Switching off the SOC term at say the Pt atoms
does not mean that there is no longer the $e_{\rm MCA}^{\rm ft,Pt}$ 
contribution because there is still a difference in the Pt-projected
density of states for the two magnetization directions due to the hybridization
of the Pt-localized orbitals with the Co-localized orbitals. And, vice
versa, when switching on the very large SOC term at the Pt atoms the 
hybridization of the Co-localized orbitals with the Pt-localized orbitals
changes drastically and this has a large effect on $e_{\rm MCA}^{\rm ft,Co}$. 
In the following we call this the ``indirect contribution'' of the Pt 
atoms to the magnetic anisotropy energy. The results of such analysis
are presented in Figs.\ 6,7. 
\begin{figure}
\begin{center}
\includegraphics[scale=0.4,angle=0]{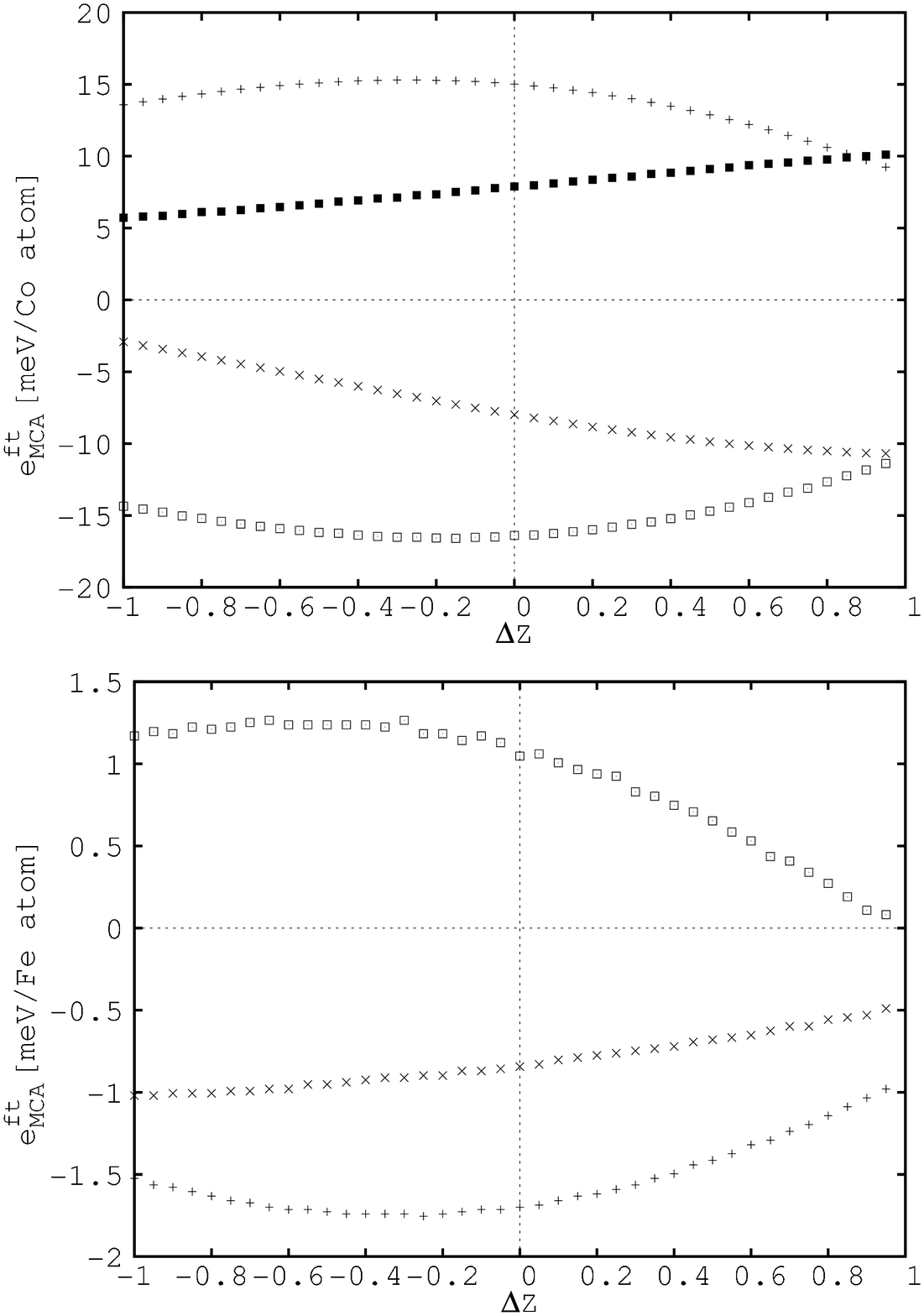}
\caption{The calculated magnetic anisotropy energy
$e_{\rm MAE}^{\rm ft}=e(\theta=0^\circ,\phi=22.85^\circ)-
e(\theta=0^\circ,\phi=-67.15^\circ)$
for the Co wire (top) and
$e_{\rm MAE}=e(\theta=0^\circ,\phi=80.40^\circ)-
e(\theta=0^\circ,\phi=-9.60^\circ)$
for the Fe wire (bottom) projected to the Co (Fe) sites
with SOC at all sites ($+$) or
SOC just at the Co (Fe) sites ($\times$), and projected to the Pt sites
with SOC at all sites ($\square$) or just at the Pt sites
($\blacksquare$).}
\end{center}
\end{figure}
\begin{figure}
\begin{center}
\includegraphics[scale=0.4,angle=0]{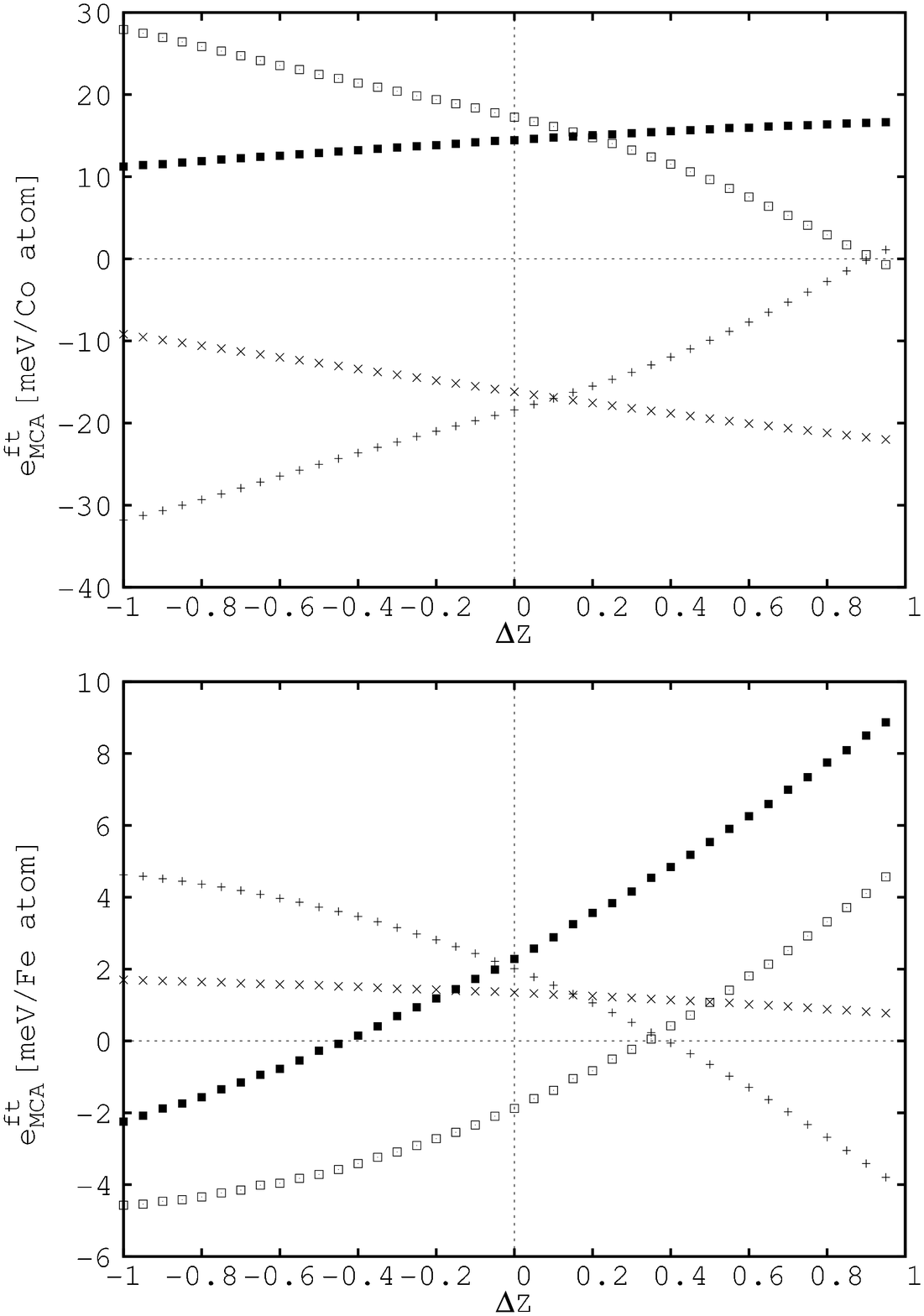}
\caption{The calculated magnetic anisotropy energy
$e_{\rm MAE}=e(\theta=0^\circ,\phi=0^\circ)-
e(\theta=90^\circ,\phi=0^\circ)$
for the Co (top) and the Fe wire (bottom) projected to the Co (Fe) sites
with SOC at all sites ($+$) or
SOC just at the Co (Fe) sites ($\times$), and projected to the Pt sites
with SOC at all sites ($\square$) or just at the Pt sites
($\blacksquare$).}
\end{center}
\end{figure}
\section{Discussion}
By looking at Fig.\ 3 it becomes immediately clear that there is a big 
influence of the Pt substrate. Whereas for the free-standing wire the
easy axis is in the wire direction, it is perpendicular to the wire
direction for the Pt-supported systems. Furthermore, there is  a big 
difference between the Pt-supported Co and Fe wires concerning the orientation
of the easy axis in the plane perpendicular to 
the wire, see Fig.\ 2. Whereas for the Fe wire the easy axis is nearly perpendicular 
to the surface normal, it is inclined by an angle
of $22.85^\circ$ from the surface normal towards the step edge in the case
of the Co wire. Remember that for the chains of Co atoms at the step
edge of Pt(997) an angle of $43^\circ$ was found 
experimentally\cite{Gambardella:2002}. The difference in the energy
between the easy and the hard axis in the plane perpendicular to the wire is about $1.3\>{\rm meV}$ per Co 
or per Fe atom, whereas the experimental value is  about $2\>{\rm meV}$
per Co atom\cite{Gambardella:2002}. In the following we want to distinguish
between the ``direct contribution'' and the ``indirect contribution'' of the Pt
atoms on the magnetic anisotropy by means of the methods (a) and (b)
of Section 2. 
\par
From Figs.\ 4,5 it becomes obvious that there is a strong ``direct contribution''
$e_{\rm MCA}^{\rm ft,Pt}$ to the total anisotropy. Over a wide range of 
the $Z$ values, the contributions of the Co(Fe) atoms and the Pt atoms are 
large but of opposite signs. For the Co wires the ``direct contributions''
from Co and Pt atoms nearly cancel each other, and the sum of the two 
contributions is close to the total anisotropy energy, indicating a small
influence of the interstitial region. For the Fe wires the two
direct contributions are much smaller than for the Co wire but the total 
anisotropy energy is of the same order of magnitude. The sum of
$e_{\rm MCA}^{\rm ft,Fe}$ and $e_{\rm MCA}^{\rm ft,Pt}$ is drastically different
form $e_{\rm MCA}^{\rm ft}$, indicating an unexpected large contribution 
of the interstitial region. \par
The plots in Figs.\ 2,3 clearly demonstrate that the magnetic anisotropy
of the Pt-supported Fe wires is strongly dominated by the SOC at the 
Pt atoms.  In the absence of this coupling, only a small magnetic anisotropy
originating from the SOC at the Fe sites remains. In contrast, to obtain the 
magnetic anisotropy of the Pt-supported Co wires, we have to take into account
the SOC for both the Pt and the Co atoms. Switching on the SOC term at the Pt
atoms modifies both $e_{\rm MCA}^{\rm ft,Co}$
($e_{\rm MCA}^{\rm ft,Fe}$) and $e_{\rm MCA}^{\rm ft,Pt}$. 
The modification of 
$e_{\rm MCA}^{\rm ft,Co}$ ($e_{\rm MCA}^{\rm ft,Fe}$) represents the 
``indirect contributions'' of Pt atoms (see (b) of section 2).
Figs.\ 6,7 represent the contribution $e_{\rm MCA}^{\rm ft,Co}$ (top) and
$e_{\rm MCA}^{\rm ft,Fe}$ (bottom) together with the respective 
contributions $e_{\rm MCA}^{\rm ft,Pt}$ on the one hand calculated 
with SOC switched on for both Co (Fe) and Pt atoms (corresponding to the 
curves shown in Figs.\ 4,5), and on the other hand calculated with SOC
on the respective other atoms switched off. 
Obviously there is a big ``indirect contribution'' of the Pt atoms, i.e., a big
change of $e_{\rm MCA}^{\rm ft,Co}$ ($e_{\rm MCA}^{\rm ft,Fe}$) when the SOC on
Pt is switched off. The ``indirect contribution'' thereby is of the same order of magnitude as the ``direct contribution''.
\par
The Pt substrate has an influence also on other magnetic properties of the
wires. For instance, we have considered a freestanding Fe biwire and Fe biwires
at the step of a Pt substrate, both for a ferromagnetic and for an
antiferromagnetic alignment of the magnetic moments on the two adjacent
monatomic wires (within each wire the alignment was ferromagnetic). The energy
difference between these two configurations was about $-20$~mRy per Fe atom for the free-standing biwire and only about $-7.7$~mRy per Fe atom for the Pt supported biwire. For comparison, the difference in energy between the ferromagnetic and the antiferromagnetic configuration in bcc~Fe at the equilibrium lattice constant of Pt is about $-25$~mRy/atom. The magnetic moment of the Fe atom for th Pt-supported biwire was very similar to the one of the Pt-supported monatomic wire.
\par
We are indebted to G.~Bihlmayer  and A.~B.~Shick for helpful discussions and
for providing their data and results. 
\bibliography{text.bib}
\end{document}